\documentstyle[11pt,aaspp4,epsf]{article}                                      
\begin{document}
\title{Model Atmospheres for Low Field Neutron Stars}
\author{Mohan Rajagopal\altaffilmark{1} and Roger W. Romani\altaffilmark{2}}
\affil{Physics Dept., Stanford University, Stanford CA  94305-4060}
\authoraddr{Physics Dept., Stanford University, Stanford CA  94305-4060}

\altaffiltext{1}{mohan@astro.stanford.edu} 
\altaffiltext{2}{Alfred P. Sloan Fellow; rwr@astro.stanford.edu} 

\begin{abstract}
	We compute model atmospheres and emergent spectra for low field ($B
\la 10^{10}$G) neutron stars, using new opacity and equation of state data
from the OPAL project. These computations, incorporating improved treatments
of flux transport and convective stability, provide spectra for hydrogen,
solar abundance and iron atmospheres. We compare our results to high field
magnetic atmospheres, available only for hydrogen.  An application to
apparently thermal flux from the low field millisecond pulsar PSR J0437--4715
shows that H atmospheres fit substantially better than Fe models.  We comment
on extension to high fields and the implication of these results for neutron
star luminosities and radii.

\end{abstract}


\section{Introduction}

Thermal radiation from the surfaces of neutron stars (NS) provides important
information about these elusive compact objects.  In particular, neutron
stars' thermal histories are already being probed by X-ray observations
({\"O}gelmann 1995); these data place useful constraints on the interior
physics and the equation of state of matter at super-nuclear densities
(Tsuruta 1995). Significant limits on heating processes such as precipitation
of magnetospheric particles and rotational energy dissipation in the crust can
also be obtained. To interpret the observed X-ray fluxes, it is often assumed
that the NS spectrum is blackbody. However in general we expect the emergent
flux to be reprocessed by the NS atmosphere, and Romani (1987) showed that the
departures from blackbody emissivities in the observed bands can be quite
substantial.

In Romani (1987) opacities from the Los Alamos Opacity Library (LAOL) (Huebner
et al. 1977) were used to generate model atmospheres and emergent
spectra for non-magnetic neutron stars with a variety of surface
compositions. More recent work (Shibanov et al. 1992; Potekhin \&
Pavlov, 1993) has treated the equation of state and opacity of pure H
atmospheres in strong ($\sim 10^{12}$G) magnetic fields. This allowed Pavlov
et al. (1995) to produce atmosphere models for high field NS;
substantial departures from blackbody spectra were found, albeit not as large
as in the non-magnetic case. Since the atomic composition of a NS surface is
quite uncertain it is also important to treat heavy element atmospheres.
Miller \& Neuhauser (1991) have calculated wave functions and energies of some
magnetized heavy elements, and Miller (1992) produced polarization averaged
bound-free cross sections and computed approximate model atmospheres.
However, separate transport of the polarization modes dramatically affects the
emergent spectrum (Pavlov et al. 1995), while at zero field bound-bound
opacity can exceed bound free by large factors (Iglesias, Rogers \& Wilson
1992), so these heavy element atmospheres may not adequately model the
emergent spectrum.  Thus since detailed treatments for elements heavier than H
in strong fields are not yet available, comparison with low-field results
remains useful.

To delimit the range over which our non-magnetic atmospheres can be applied,
we must estimate the field required to perturb the opacity significantly at
frequencies of interest.  In the presence of strong magnetic fields, the
free-free opacities for one of two polarization modes perpendicular to the
field and for radiation parallel to the field are strongly suppressed.  For
Thompson scattering, the cross section decrease occurs below the electron
cyclotron energy, i.e.  $E_\gamma < h\nu_B = 11.6 B_{12}\;$keV, where $B =
10^{12} B_{12}$G (Canuto, Lodenquai \& Ruderman, 1971).  For inverse
Bremsstrahlung suppression, both $E_\gamma$ and $kT$ should be less than
$h\nu_B$ (Pavlov \& Panov, 1976); at typical NS temperatures of $\sim 10^6$K,
these effects would affect the ROSAT PSPC band (0.1-2.4keV) when $B_{12} >
0.01$.  Magnetic effects on the bound-free and bound-bound opacities depend on
the full quantum states of magnetized atoms, but some generalizations are
possible.  In the case of hydrogen, detailed calculations are available
(e.g. R\"{o}sner et al. 1984); for the field $B_0 = 1.2\times 10^{9}$G, at
which $h\nu_B = 1{\rm Ry}$, these authors find the ground state binding energy
$E_B$ is increased by $\sim 40\%$.  Thus this bound-free feature too is
altered when $h\nu_B \sim E_\gamma$.  The effect vanishes rapidly at lower
fields, while for large fields we find their results give $E_B \sim
(2.5B/B_0)^{1/3}\;{\rm Ry}$.  To estimate the minimum perturbing field in
general, consider the magnetic contribution to the Hamiltonian of a hydrogenic
atom (R\"{o}sner et al. 1984):
\begin{equation}
H_{\rm mag}/1{\rm Ry} \equiv H_1 + H_2 = 2\beta (m\pm 1) + \beta^2 r_\perp^2
\end{equation}
where $\beta = B/(4 B_0)$, $\;m$ is the quantum number for angular momentum in
the field direction, $\pm1$ is for the spin, and $r_\perp$ (in units of Bohr
radii) is the atomic radius perpendicular to $B$.  Here $H_1$ is the Zeeman
effect, while $H_2$ is sometimes referred to as the quadratic Zeeman effect.
For an energy level $E_B$ with principal quantum number $n_p$ and a nucleus
with charge $+Z$, we define the critical field $\beta_c$ to be that which
gives $H_{\rm mag} \sim E_B$.  Using $r_\perp \sim n_p^2/Z^2$ and $E_B \sim
Z^2/n_p^2\,{\rm Ry}$ we find that $H_1 \sim E_B \Longrightarrow \beta \sim
(Z/n_p)^2$ (for $m\pm 1$ of order unity), and $H_2 \sim E_B
\Longrightarrow \beta \sim (Z/n_p)^3$. When $n_p > Z$, $H_2 > H_1$ and
so the critical field scales as $n_p^{-3}$. This scaling is borne out by the
results of R\"{o}sner et al. (1984) for excited states of Hydrogen. However,
in NS atmosphere conditions only modest excitation of a species is obtained
before it is further ionized, so for larger $Z$ the perturbation $H_1$
dominates in almost all cases.  Its dependence implies that spectral features
will be perturbed when $h\nu_B \sim E_\gamma$, in agreement with Miller \&
Neuhauser (1991), who find for hydrogenic atoms of atomic number Z that the
lowest energy levels follow $E_Z(B) = Z^2E_H(B/Z^2)$, where $E_H$ is the
corresponding hydrogen energy level at field $B/Z^2$.  They also find the
helium ground state to follow this scaling, with a screened $Z_{\rm eff}$.
Line positions, which represent a difference between two energy states, will
deviate substantially from their zero field values and ordering even before
$h\nu_B$ reaches either energy, but a shell feature such as the iron L-edge at
0.7 keV in our spectra should persist to $B_{12}\sim 0.7/11.6 = 0.06$.  Though
the atmospheric structure will be affected for $h\nu_B > kT$ as the peak flux
is altered, gross opacity features above the thermal peak will persevere to
higher fields. Our non-magnetic atmospheres are therefore entirely suitable
for low-field ($10^8-10^{10}\;$G) NS such as millisecond pulsars, and
indicative of some spectral features at higher fields.

Two developments suggest that a re-examination of the heavy element,
non-magnetic NS atmospheres is timely.  First, improved treatments of
astrophysical opacity using the OPAL code (Iglesias et al. 1992; and
references therein) and by the OP collaboration (Seaton et al. 1994)
have shown that the LAOL results substantially underestimate the opacity,
especially for iron-group elements in the $\sim 10^5$K temperature range.  Use
of the new opacities already seems to resolve several puzzles in stellar
astrophysics, including discrepancies between observations of Cepheid
pulsations and models based on LAOL data (see Iglesias et al. 1992).
The improvements to the opacity and equation of state (EOS) are particularly
important to the NS spectrum problem, as the photosphere in these atmospheres
forms precisely at the densities and temperatures where the changes to the
LAOL results are largest.  Secondly, recent observations have resulted in a
number of strong (ROSAT) and potential (EUVE) detections of apparently thermal
emission from nearby neutron stars, providing limited spectral information in
the EUV/soft X-ray band. In particular, some low field (millisecond) pulsars
have now been detected. It is clear that the next generation of space
facilities in this band will provide significant constraints on NS thermal
compositions and emissivities.

Thus to provide an improved baseline of models that illustrate spectral
changes with surface composition and are directly applicable to low field
neutron stars, we have constructed neutron star atmospheres and emergent
spectra using the new opacities and consistent EOS, for pure iron, pure
hydrogen, and the solar abundances of Grevesse \& Noels (1993).  Improvements
in the radiative transfer are also introduced and the resulting spectra should
superceed those of Romani (1987). Initial results of these computations have
been reported in Romani, Rajagopal, Rogers \& Iglesias (1995).  In Section II
we describe the calculation of the model atmosphere structures; in Section III
we discuss convectional stability of the atmospheres, and present emergent
spectra.  Spectra for various compositions are compared to black bodies and
the high field hydrogen atmospheres of Pavlov et al. (1995).  In Section
IV we provide an initial application, fitting the spectra to ROSAT
observations of the millisecond pulsar J0437--4715. Section V describes
implications of these results and prospects for future work.

\section{Model Atmospheres}

\subsection{OPAL Data}

	Because the EOS treatment allows extension to high density, OPAL EOS
and opacity tables (kindly supplied by F. Rogers and C. Iglesias) are our
primary data source for these models.  OPAL opacity computations are based on
the method of detailed configuration accounting using LS coupling (Iglesias,
Rogers, \& Wilson, 1987), and now use full intermediate coupling for iron to
incorporate spin-orbit interactions (Iglesias et al. 1992).  Abundances
of all possible ions are obtained from an activity expansion of the grand
canonical ensemble (Rogers 1986), and tables of the equation of state (EOS)
completely consistent with that used for the opacity calculations are
available (Rogers, Swenson, \& Iglesias, 1995).  For the present models, the
EOS tables provide pressure and other thermodynamic variables as a function of
temperature and density, interpolated using routines supplied with the data.

As an example, for pure iron the opacity grid was obtained for each decade of
$R \equiv \rho/T_6^3$, where $T_6 = T/(10^6\;K)$, from -5 to +1; with ${\rm
Log(}T)$ from 4.5 to 7.5 every 0.25, and with $10^4$ photon energies linearly
spaced over $0.002 < u \equiv E_\gamma/(k_B T) < 20.0$.  As needed, we
interpolate $\log \kappa$ linearly against $\log R$ and $\log T$ and linearly
against $u$ (at constant $\rho$ and $E_\gamma$).  To allow extension to high
photon energies, we use additional tables for the same $R$ and $T$ spaced
linearly in $E_\gamma$ from 1eV to 10keV at 1eV intervals.  These tables, used
only when $E_\gamma > 20k_BT$, allow us to solve the transfer equation for
$E_\gamma$ up to 10 keV, through the lower temperatures present in the outer
atmospheres.  The data for the pure H and solar (Grevesse \& Noels 1993)
abundances are similar, densely covering the thermal peak with extension up to
1 keV.

\subsection{Model Atmosphere Construction}

We define {\it ab initio} a vector of 120 mean optical depths $\tau_{T}$,
logarithmically spaced between $10^{-3}$ and $10^{3.5}$.  These correspond to
physical depths through ${\rm d}z={\rm d}\tau_T/\rho\kappa_T$, where
$\kappa_T$ is the total mean opacity in ${\rm cm}^2{\rm g}^{-1}$ including
electron conductivity (more below).  The temperature and density at these grid
points are adjusted until both hydrostatic equilibrium and energy steady state
(total astrophysical flux $F =\sigma T_{\rm eff}^4/\pi$) are achieved at all
depths.  For each of the three compositions, we generate six model atmospheres
with $T_{\rm eff}$ from $10^{5.25}$ to $10^{6.5}$, four per decade.  For the
light element mixtures, the EOS tables stop before $\tau_T = 10^{3.5}$ in the
two coolest atmospheres, and for iron in the one coolest and two hottest. In
these cases, the depth grid extends to $\tau_T = 10^{2.5}$; this affects only
the highest $E_\gamma$ slightly.

At each frequency $\nu$, the outward flux $F_\nu$ at the frequency-specific
optical depth $\tau_\nu$ is obtained from the Milne integral
\begin{equation}\label{transfer}
F_\nu(\tau_\nu)\; =\; 2 \int_{\tau_\nu}^\infty S_\nu(\tau'_\nu)E_2
(\tau'_\nu-\tau_\nu){\rm d}\tau'_\nu \;-\; 2 \int_0^{\tau_\nu} 
S_\nu(\tau'_\nu) E_2(\tau_\nu-\tau'_\nu){\rm d}\tau'_\nu
\end{equation}
where $S_\nu$ is the radiation source function, and $E_2$ is the second
exponential integral, which has maximum of one at zero, with decay length of
order unity (Mihalas 1978).  Similarly, the intensity $J_\nu$ is given by
\begin{equation}\label{J_trans}
J_\nu(\tau_\nu) = \frac{1}{2} \int_0^\infty S_\nu(\tau'_\nu) E_1 
\;\vert \tau'_\nu - \tau_\nu \vert\; {\rm d}\tau'_\nu
\end{equation}
where $E_1$ is the first exponential integral.  We assume Local Thermodynamic
Equilibrium, taking the source of radiation at each depth to be a Planck
function at the local temperature $B_\nu(T)$.  The atmosphere models begin
with the grey opacity solution $T^4(\tau)=(3/4)T_{\rm eff}^4(\tau + q(\tau))$,
using tabulated values for $q(\tau)$.  In general, in a non-grey purely
radiative atmosphere, at depths so large that radiative flux is given at all
frequencies by the diffusion approximation
\begin{equation}\label{diffusion}
F_{\nu} = -\frac{4}{3}\left( \frac{1}{\kappa_{\nu}} \frac{\partial B_{\nu}}
{\partial T} \right) \left(\frac{{\rm d}T}{{\rm d}z}\right)\;,\;\;\;\;\;\;\;
{\rm giving}\;\;F_{\rm rad} = \int F_\nu {\rm d}\nu = -\frac{16\sigma}
{3\rho\kappa_R} T^3 \frac{{\rm d}T}{{\rm d}z}\;,
\end{equation}
the grey solution is obtained on the optical depth scale defined by the
Rosseland mean opacity. In the dense atmosphere of a neutron star, however,
energy transport by electron conduction can be significant.  We thus define an
equivalent conduction opacity $\kappa_c$ by analogy with equation
(\ref{diffusion}) so that $-\lambda_c\;{\rm d}T/{\rm d}z = F_{\rm cond} \equiv
(16\sigma T^3) /(3\rho\kappa_c)\;{\rm d}T/{\rm d}z$, where $\lambda_c$ is the
thermal conductivity. Here we use $\kappa_c = (2.5\times 10^4/N_e)\; (Z^2/A)\;
(T_7^{1/2}/\rho)\;\; {\rm cm}^2/{\rm g}$ where $N_e$ is the number of free
electrons per AMU, $A$ is the average atomic mass, $Z = A N_e$ the average
ionic charge, $T_7$ is in $10^7$ K, and $\rho$ is in ${\rm g/cm}^3$. Adding
harmonically with the radiative opacity provides a total mean opacity
$\kappa_T^{-1} = \kappa_R^{-1} + \kappa_c^{-1}$ on whose depth scale the
initial grey model provides the correct atmosphere solution at large $\tau$.

We start with the grey atmosphere temperature structure, and impose
hydrostatic equilibrium starting at the surface ($P=0$).  For each $\tau_T$
layer, using an initial density $\rho_0$ from the layer above, we calculate
the pressure at the base of that layer from ${\rm d}P/{\rm d}\tau_R =
g/\kappa_R(\rho_0,T)$.  We then obtain a corrected $\rho_1 (T_0, P_0)$ from
the EOS tables, and iterate to convergence.  At each iteration, we extract the
opacity on an 800 bin logarithmic frequency grid extending to 10keV, saving
harmonic means of 50 OPAL opacities sampled within the bin. This grid of
$\kappa_\nu$ values is used to produce the Rosseland mean opacity $\kappa_R$.
Once convergence is reached at each depth, the final $\kappa_\nu$ and
$\kappa_R$ are saved, and similar arithmetic bin averages used to produce the
Planck mean $\kappa_P$.  Once hydrostatic equilibrium is achieved for the
whole atmosphere, we solve the transfer equation (\ref{transfer}) for each of
the same 800 frequencies, first computing $\tau_\nu$ corresponding to each of
the 120 $\tau_{T}$ by integrating ${\rm d}\tau_\nu = (\kappa_\nu/\kappa_R)
{\rm d}\tau_R$ down from the surface.  At very large depth, where
$S_\nu(\tau_\nu)$ varies slowly, we are in the diffusion limit and $F_{\nu}$
is obtained from equation (\ref{diffusion}); the intensity $J_\nu$ always
comes directly from equation (\ref{J_trans}).  Total flux $F$ and intensity
$J$ are obtained by summing over the frequency grid, while the flux and
absorption mean opacities $\kappa_F$ and $\kappa_J$ are obtained by
appropriately weighted sums over the $\kappa_\nu$.

The temperature structure is then corrected using the Lucy-Uns\"{o}ld
procedure to drive towards the desired steady state flux:
\begin{equation}\label{lucyu}
\Delta T(\tau) = \frac{1}{16\sigma T(\tau)^3} \left[ \frac{\kappa_J}{\kappa_P}
\left( 3 \int_0^\tau \frac{\kappa_F(\tau')}{\kappa_R(\tau')} \Delta F(\tau') 
{\rm d}\tau' + 2 \Delta F(0) \right) - \frac{\kappa_R}{\kappa_P}
\frac{{\rm d}\Delta F(\tau)}{{\rm d} \tau_R} \right] ,
\end{equation}
(Mihalas, 1978) where $\tau \equiv \tau_T$ as defined above, and $\Delta F =
\sigma T_{\rm eff}^4/\pi - F_{\rm rad} - F_{\rm cond}$.  The corrected 
temperature run is then smoothed and any temperature inversions in the outer
atmosphere are removed. The procedure is iterated to convergence $(\pi\Delta
F/\sigma T_{\rm eff}^4 < 0.005$ at all depths).

\goodbreak
\section{Results}

\begin{figure}[htb]
\plotone{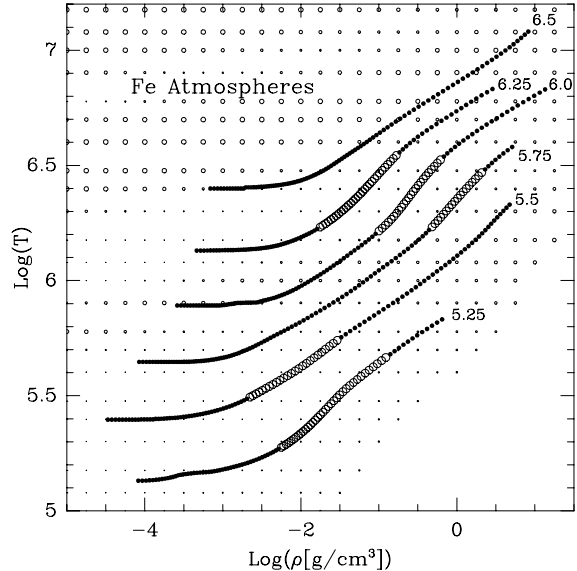}
\caption{Iron atmosphere structures.  Open circle regions of the curves are
convectively unstable, and size of circles in the field indicates the 
adiabatic index of the gas.}
\end{figure}

We show the converged temperature vs. density runs for iron in Figure 1.  Open
circles indicate regions of convective instability, by definition where 
$({\rm d} \log T/{\rm d}\log P)_{\rm atm} \equiv \Delta_{\rm atm} > 
\Delta_ {\rm ad} \equiv ({\rm d}\log T/{\rm d}\log P)_{\rm adiabatic}$.  
The value of $\Delta_{\rm ad}$ for iron from the EOS is indicated by the
size of the background circles; instability in the radiative/conductive
solutions occurs chiefly where it is low, in ionization zones.  To gauge the
importance of this instability we compute the energy gain ratio of a rising,
but radiating, bubble
\begin{equation}\label{egain}
\Gamma \approx {{\kappa \rho^2 c_P} \over {24\sigma T^3}}
l_{\rm mfp} v
\end{equation}
({\it e.g.} B{\"o}hm-Vitense, 1992) where $v$ is the mean bubble speed and we
take the bubble mean free path $l_{\rm mfp}$ to be the atmospheric scale
height $P\,{\rm d}z/{\rm d}P$.  For small $\Gamma$, departures from the
radiative temperature gradient scale as $9\Gamma^2/4$. If the bubbles rise
freely and quasi-adiabatically, then equating the potential energy gain to the
kinetic energy gives $v \sim [g ({\rm d}T/{\rm d}z|_{\rm rad} - {\rm d}T/{\rm
d}z|_{\rm ad})/T]^{1/2} l_{\rm mfp}$ as an upper limit to the bubble velocity.
In this worst case picture, $\Gamma$ is as large as $\sim 5$ at large depth in
the ${\rm Log}T = 5.75$ atmosphere.  Accordingly the emergent flux at several
keV, where $\tau_\nu$ is modest at large depth, will be slightly affected.
However, even the lowest known magnetic fields for neutron stars are $\sim
10^8$G. If $B^2/8\pi > \rho g ({\rm d}T/{\rm d}z|_{\rm rad} - {\rm d}T/{\rm
d}z|_{\rm ad})\,l_{\rm mfp}^2/T$, which is true for $B > 10^7\;$G throughout
all the atmospheres we have computed, then the magnetic field will suppress
the convection.  Mass motions can then only occur on the diffusion time scale,
which for a partly ionized plasma allows $v_{\rm diff} \sim 1.1 \times 10^4
T_6^{-3/2} (1{\rm cm}/l_{\rm mfp})\, {\rm cm\, s^{-1}}$.  Under these
conditions $\Gamma$ is no larger than 0.04 except in the coldest iron
atmosphere, in which convection might cause departure from the radiative
temperature gradient at the 10\% level.  Accordingly, even when the magnetic
field is too small to affect the EOS and opacities it should strongly suppress
convection in most cases.

\begin{figure}[htb]
\plottwo{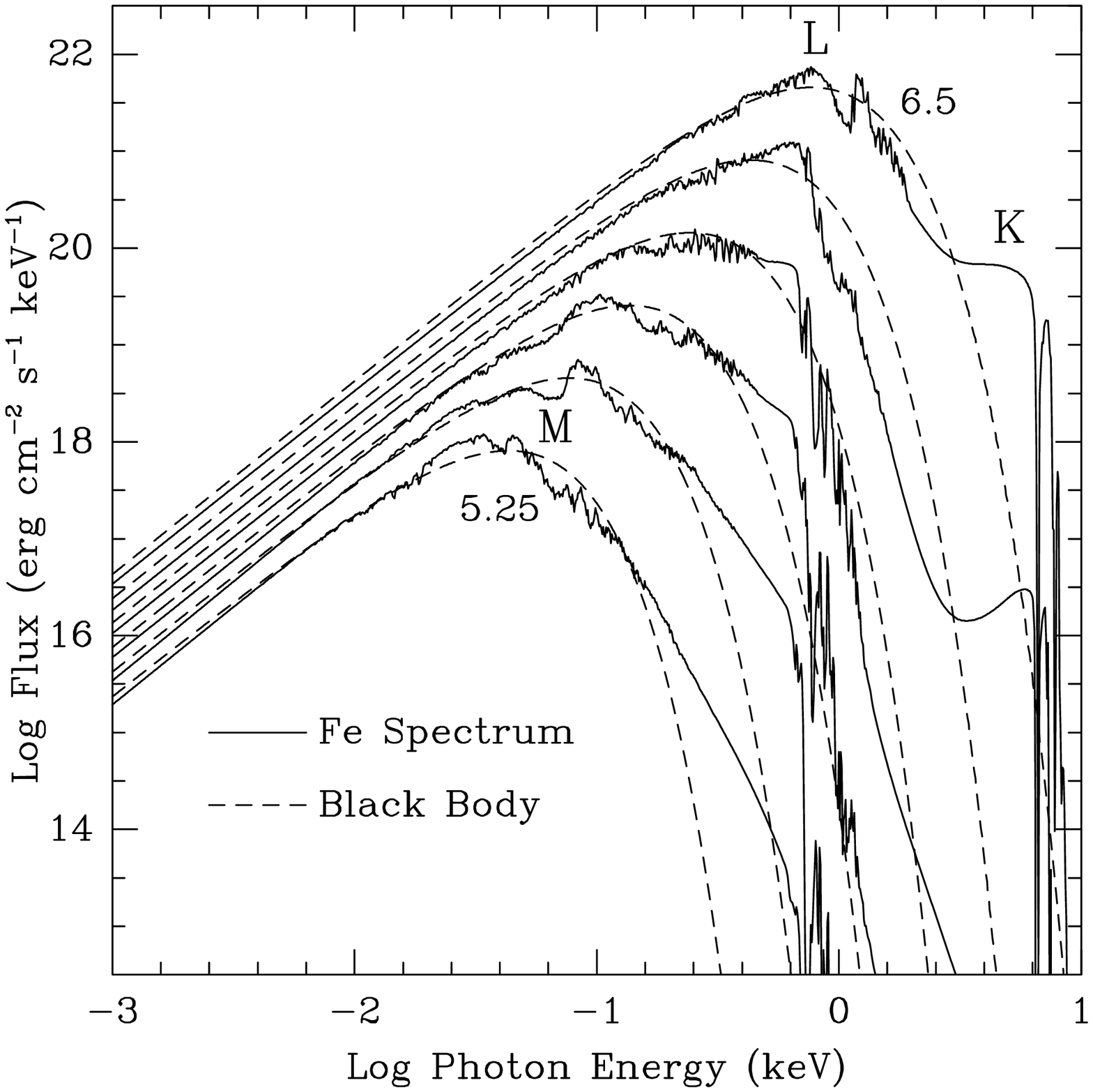}{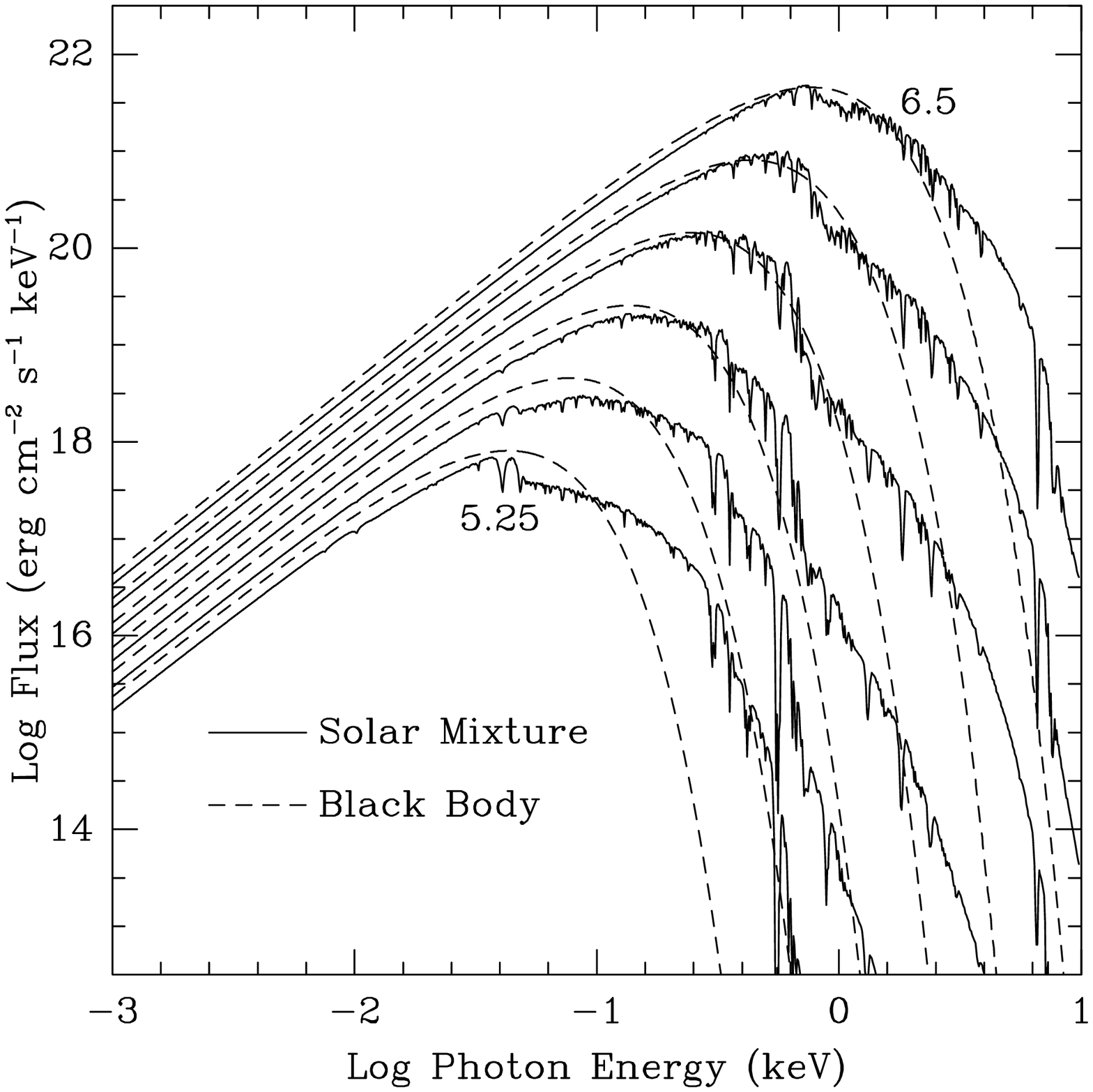}
\caption{Emergent spectra for atmospheres containing heavy elements.  See text
for comments.}
\end{figure}

We calculate emergent spectra by solving the transfer equation at the edge of
the atmosphere.  To resolve lines from intermediate depths, we compute
emergent spectra on a $2 \times 10^4$ point energy grid, logarithmically
spaced between 1eV and 10keV.  The spectra, binned down to $10^3$ energies,
are shown in Figure 2.  Much coarser binning still is appropriate to the
resolution of current NS observations, and as argued above, broad-band
features will be stable even to moderate NS fields.  The iron spectra all show
features at the K, L and M edges, which are also present in the solar
abundance spectra.  Narrow absorption lines are formed at small physical
depth, while strong pressure broadening affects lines from the deep layers.
All three compositions are compared in Figure 3 (left); the hydrogen spectra
show the expected hardening due to the $\nu^{-3}$ dependence of the opacity,
which allows flux to shift above the Wien peak.

The two major spectral effects of magnetic field on hydrogen are visible in
Figure 3 (right).  First, the high energy hardening is not nearly as marked,
since the diminished magnetic opacities which dominate transfer fall much less
steeply than $\nu^{-3}$ (Shibanov et al. 1992, Fig. 1).  Secondly, the higher
photoionization threshold increases absorption between $\sim$ 0.1 and 0.3 keV
for the fields shown; the sharp onset is smoothed out by pressure effects on
the atomic initial state (Pavlov et al. 1995).  At the lower field, this
effect causes the prominent dip in the $\log T = 5.5$ spectrum; at $\log T =
6.0$ the dip is minimal, due to the lesser abundance of atomic hydrogen.  At
the higher field, only the onset is seen in the $\log T = 5.5$ spectrum, and
it is clear this edge could dominate a fit.

\begin{figure}[htb]
\plottwo{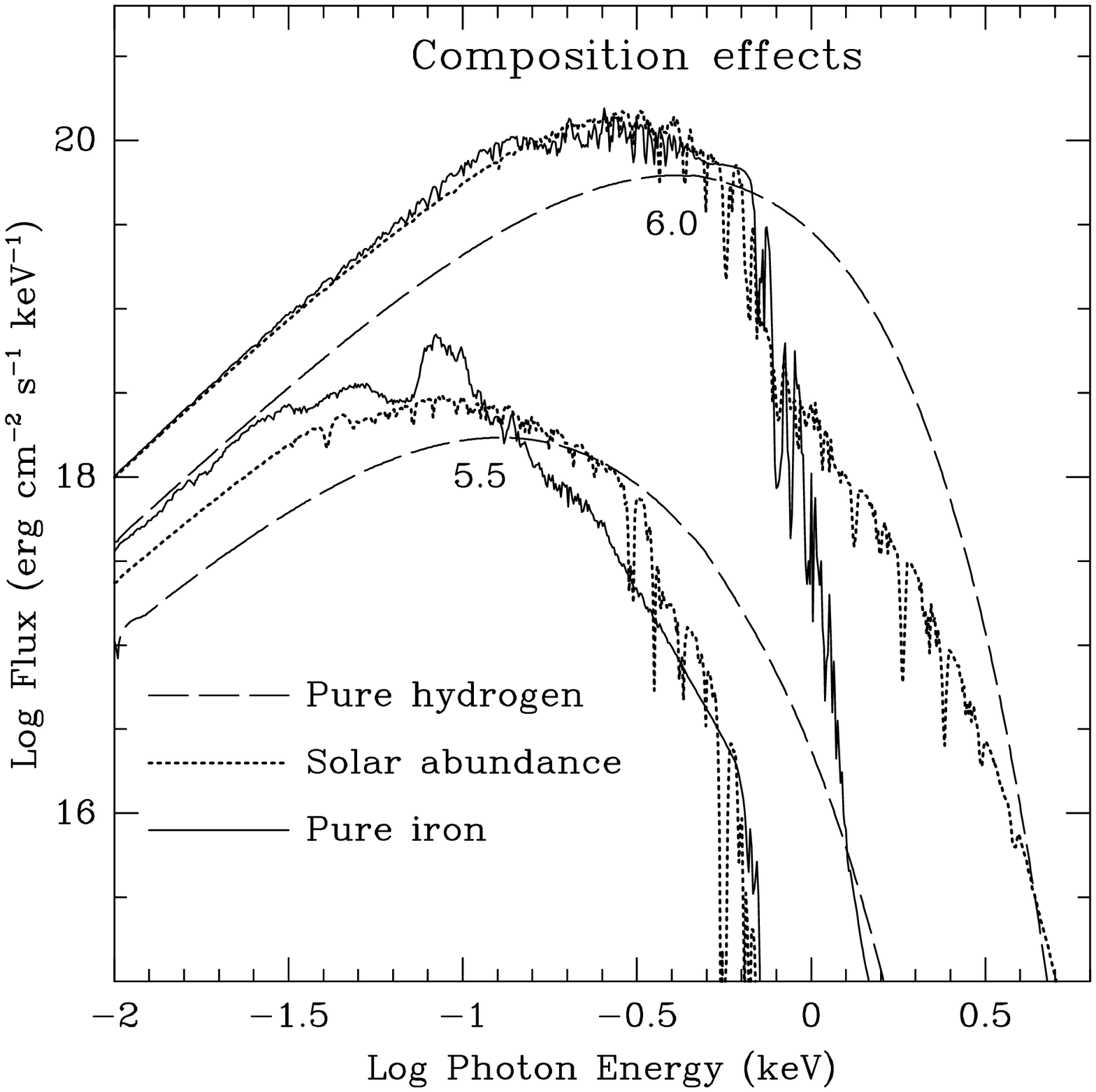}{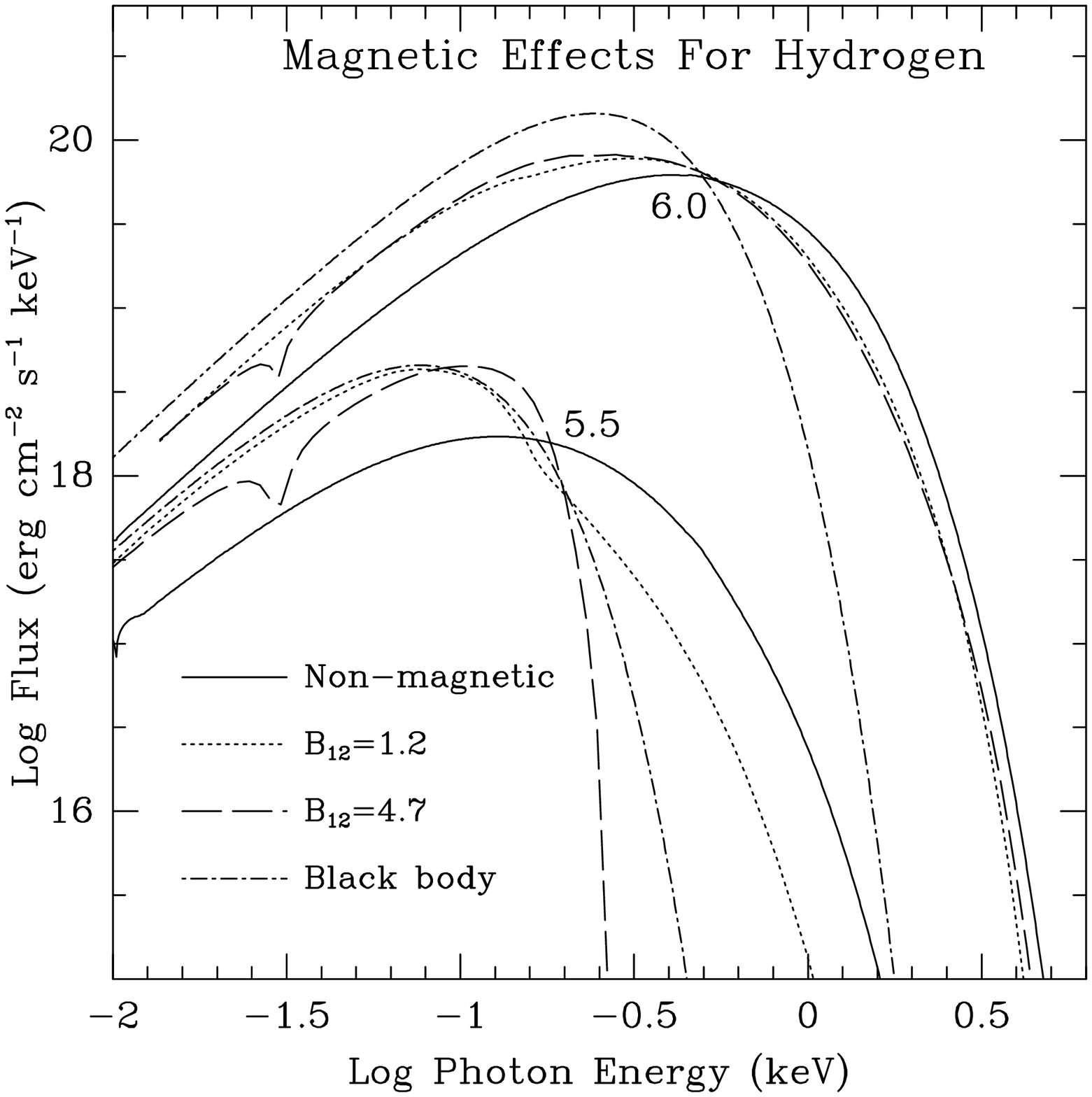}
\caption{Left --  comparison of different atmospheric compositions at two 
$T_{eff}$.  Right --  comparison of hydrogen spectra from the present work 
with the magnetic atmospheres of Pavlov et al. (1995).}
\end{figure}

\section{Application}

To apply these spectral results to soft X-ray data, we convert our spectra for
each composition into a tabulated model for the XSPEC package.  Our model
supplies unredshifted flux at the NS surface.  These spectra are subjected to a
specified gravitational redshift, and overall normalization is fit as a free
parameter.  Model surface flux $F_{\rm NS}$ is related to model observed flux
$F_{\rm obs}$ by:
\begin{equation}\label{fluxnorm}
F_{\rm obs}(E_\gamma) = \frac{F_{\rm NS}[(1+Z)E_\gamma]}{(1+Z)} \left( 
\frac{A_{\rm em}}{\Omega D^2} \right) \left( \frac{1}{1+Z} \right),
\end{equation}
where area $A_{\rm em}$ of the neutron star at distance $D$ emits into solid
angle $\Omega$; $Z$ is the gravitational redshift, and the last factor due to
time dilation.  For full surface emission $\Omega = 4\pi$; if the star has
a small emission region such as a heated polar cap we take $\Omega = 2\pi$, 
noting that
limb-darkening would tend to decrease this value while gravitational 
light bending near the NS surface would cause it to increase.

\begin{figure}[htb]
\plotone{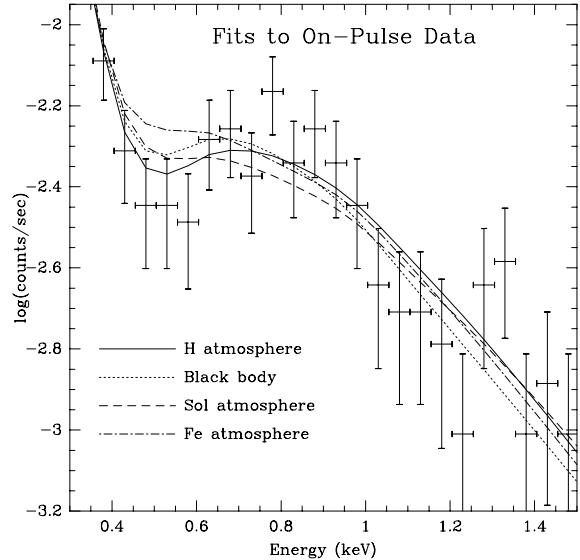}
\caption{Best fits of model thermal components plus off pulse power law, to 
J0437--4715 ROSAT PSPC data in a crucial energy range.  The model fits to the 
high-count soft peak below 0.3 keV are indistinguishable.  Pi channels of the 
PSPC detector are here binned in groups of five, for display purposes only.}
\end{figure}
 
As an example, we fit our models and magnetic hydrogen atmosphere emergent
spectra for $1.2 \times 10^{12}\;$G and $4.7 \times 10^{12}\;$G, (Pavlov {\it
et al.}, 1995), to a hypothesized thermal component in the X-ray emission of
the nearby millisecond (P = 5.7 ms) pulsar J0437--4715 measured by ROSAT.  Due
to its low inferred surface dipole field of $8.3 \times 10^8$G this is an
ideal candidate, though its large spindown age $\sim 7.6 \times 10^8\;$yr
(Johnston et al. 1993) suggests that any thermal emission should come from
reheating.  We have re-analyzed the September 20-21 1992 PSPC data studied by
Becker \& Trumper (1993).  This data set is in three segments, each of about
2000 seconds' duration, the last two about one hour apart and separated from
the first by about one day.  We use 1300 photons from a circle 2 arcmin in
radius centered at the X-ray position.  Arrival times at the spacecraft clock
are barycentred using standard IRAF PROS routines, and corrected for the
pulsar's orbital position using an updated radio ephemeris (Johnston et
al. 1993; Bell, 1995). The nominal phase of the X-ray pulse in the last two
segments aligns with the radio peak to within the ROSAT clock accuracy.  An
apparent ROSAT clock error after the first data segment required us to phase
it separately, and make the necessary shift to align it with the other two.
Based on the resulting light curve, we divide the data set into on- and
off-pulse phases of equal duration; the on-pulse interval contains $\sim 60\%$
of the counts.  While the clock error decreases confidence in absolute phasing
somewhat, the measured alignment of the X-ray pulse with the radio peak
(inferred from polarization data to be coincident with the neutron star
magnetic axis; Manchester \& Johnston 1995) supports the picture that this
pulse represents thermal emission from a reheated polar cap.  Hypothesizing
that cap emission is superimposed on an unpulsed background magnetospheric or
plerionic in origin, we fit a power law to the off-pulse data, finding a power
law index of 2.5, with absorption column density of $1.1
\times 10^{20}\;{\rm cm}^{-2}$, in agreement with the results of Becker \&
Tr\"{u}mper (1993).  We then added each thermal candidate model to the frozen
power law with absorption held fixed, finding the best-fit parameters to the
on-pulse data for blackbody spectra, atmosphere spectra for our three
compositions, and spectra from the magnetic hydrogen atmospheres.  All models
are processed through the current PSPC response matrix for comparison to the
pi-channel data, and given $Z=0.306$ redshift for a standard $1.4{\rm
M}_\odot$ neutron star 10km in radius.  Fit parameters are determined via the
maximum likelihood method using the C-statistic on unbinned data; the $\chi^2$
for these fits are then determined from data binned to at least 20 counts per
bin, with bin-width at least 1/3 the half-max width of the ROSAT energy
response (Table 1).  There are 12 DOF for the magnetic models, 11 for the
others.  Figure 4 shows the best fit models for black body and for each
composition, smoothed through the detector response.  Given the fitted
normalization of the thermal component, we use equation (\ref{fluxnorm}), the
radio dispersion pulsar distance of 140pc, $Z=0.306$ as above, and $\Omega =
2\pi$ to infer the emission areas (Table 1).

\begin{table}[htb]
\begin{center}
Table 1:  Fits to PSR J0437--4715 PSPC data \vspace{5 mm}

\begin{tabular}{|c|c|c|c|c|} \hline \hline
       Model&$T_{\rm eff}(10^6$K)&$A_{\rm eff}({\rm km^2}$) & 
            $\chi^2/{\rm DOF}$ & $P(>\chi^2)$ \\[4pt] \hline
       BB & $1.80$ & $0.31$ & 1.43 & .205 \\
       H & $0.60$ & $22$ & 0.95 & .605 \\
       Fe & $1.82$ & $0.34$ & 2.5 & .002 \\
       Sol & $1.4$ & $0.73$ & 1.5 & .166 \\
       Mag(1.2) & 1.0 & 4.3 & 0.91 & .642 \\
       Mag(4.7) & 1.0 & 4.2 & 0.93 & .623 \\ \hline
\end{tabular}
\end{center}
\end{table}

The fit for both atmospheres containing iron is significantly poorer because
the strong L-edge (see Fig. 2) provides excess flux near 0.7 keV where the
data show a deficit. Further, the model counts fall sharply above the edge
where the data show an excess.  The problem is exacerbated by redshifting the
model edge to 0.6 or 0.5 keV.  The BB and H fits are acceptable.  Following
our hypothesis that the thermal emission comes from re-heated polar caps, we
can check the fitted effective areas against the cap area expected in the
aligned dipole rotator model: the polar cap radius is $r_{\rm pc} =
r\sin^{-1}(r/r_{\rm lc})^{1/2}$, where r is the pulsar radius and $r_{\rm lc}
= (Pc/2\pi)$ the light cylinder radius.  In the case of J0437--4715, for $r =
10\;{\rm km}$ we have $r_{\rm pc} = 1.9\;{\rm km}$, giving $A \sim \pi r_{\rm
pc}^2 = 12\;{\rm km}^2$.  Only the H atmosphere fits give areas close to this
standard polar cap; BB and Fe fits imply areas $\sim 40$ times too small.
While the magnetic H fits and areas are acceptable, the $4 \times 10^4$G light
cylinder field inferred from spindown torques would not allow surface $ B\sim
10^{12}$G unless the magnetic structure were of octupole order or higher.  The
resulting cap structure and thermal pulse would differ greatly from the dipole
values inferred for PSR J0437--4715.

\section{Conclusion}

The behavior of NS spectra at energies high in the ROSAT band is crucial for
temperature determination, as interstellar absorption often leaves only the
Wien-like tail detectable.  In that case, the inferred temperature depends
strongly on the composition model adopted: in most cases the atmospheric
temperatures are substantially lower than those inferred for blackbodies,
although high temperature Fe atmospheres may require a $T_{\rm eff}$ slightly
higher than black body.  Even with present limited spectral information and
moderate absorption this composition dependence is significant.  As an
illustration, we have drawn sample ROSAT PSPC data sets as seen with
absorption of $10^{20}\;{\rm cm}^{-2}$ from our redshifted ($Z=0.306$) models,
and fit them with black body spectra.  Figure 5 shows the ratios of fitted to
actual temperature.  For the $B_{12} = 4.7$ hydrogen atmosphere at it's lowest
two temperatures, the black body temperature is driven down by the
photoabsorption edge (above, and Figure 3).  Standard cooling curves for
neutron stars (e.g. modified URCA processes) run about four times hotter than
curves for more exotic processes (e.g. direct URCA, pion condensates) before
photon cooling takes over (\"{O}gelman, 1995; Tsuruta, 1995) at $t \sim
10^6$y.  Black body fits to thermal spectra from
\"{O}gelman's four initial cooling candidates place them quite near the
standard curve, but Figure 5 demonstrates that even magnetic light element
atmospheres could move their correct positions close enough to the cooler
exotic process cooling to confuse matters considerably.

\begin{figure}[htb]
\plotone{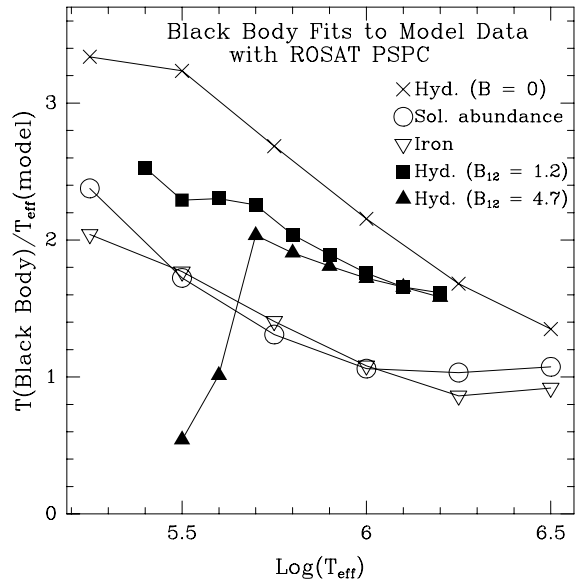}
\caption{Fractional temperature error incurred when simulated 3000-count 
ROSAT PSPC observations of our model spectra are fitted with a black body.}
\end{figure}

Our emergent spectra are available in electronic form for the reader
interested in determining flux implications for a particular band or detector,
either as text files or as XSPEC models.  For the pulsar J0437--4715, enough
spectral information exists to break the composition/$T_{\rm eff}$ degeneracy,
favoring a pure hydrogen atmosphere quite strongly.  The implied polar cap
area is in much better agreement with theoretical expectation than that
inferred from a blackbody fit. Future missions (e.g. AXAF \& XMM) should be
able to make similar measurements of other millisecond pulsars.
Interestingly, non-magnetic fits to the `thermal' emission of young,
high-field pulsars seem in many cases to favor H atmosphere models to
blackbody spectra (\"{O}gelman 1995 and references therein). While this agrees
with the the compositional inference from PSR J0437--4715, comparison with
heavy element magnetic atmospheres is needed to substantiate this result.  A
correct treatment of the detailed spectrum of such high $B$ neutron star
atmospheres will require extensive improvements to present atomic structure
and opacity computations, although approximate models following broad-band
features may now be feasible (work in progress).  For younger pulsars, such
models together with data from the next generation of X-ray satellites should
enable a serious study of NS surface conditions, including composition,
magnetic field and even redshift, with important implications for NS evolution
and the EOS of matter at very high densities.

\bigskip
\bigskip

We are indebted to Forrest Rogers and Carlos Iglesias for computation of the
OPAL data for the Fe models and assistance with their interpretation.  We also
thank G. G. Pavlov for providing high-field hydrogen spectra, 
I.-A. Yadigaroglu for assistance with the J0437 data, and the referee for
a careful reading.  RWR was supported in
part by an Alfred P. Sloan fellowship and NASA grant NAGW-2963, and MR in part
by a fellowship from the National Science and Engineering Research Council of
Canada.  This research has made use of data obtained through the High Energy
Astrophysics Science Archive Research Center Online Service, provided by the
NASA-Goddard Space Flight Center.

\end{document}